# On a discussion about the determination of surface characteristics of microemulsion droplets from static and dynamic scattering experiments

V. Lisy[*]

*Institute of Physics, P.J. Safarik University, Jesenna 5, 041 54 Kosice, Slovakia*

___

The interpretation of static and dynamic scattering experiments on droplet microemulsions is discussed. We review the arguments given in our previous papers that call into question the methods of current determination of basic phenomenological characteristics of the droplet surfactant monolayer, such as the bending and Gaussian rigidities. We give also responses to our criticism by T. Hellweg, B. Farago, M. Gradzielski, D. Langevin, and S. Safran (Colloids and Surfaces A 221 (2003) 257 and some other works by these authors). They agree with some of our objections; the rest of their response is shown to be flawed. We also show that in many points their way of discussion is improper and misleading. It is concluded that the values of the parameters of microemulsion droplets, as extracted in their experiments, are not reliable and should be reexamined.

___

In the frame of the widely accepted Helfrich's phenomenology of interfacial elasticity the properties of droplet microemulsions are described using a few parameters of the surfactant monolayer of the droplets. These characteristics, such as the bending rigidity $\kappa$ and the Gaussian modulus $\bar{\kappa}$, are thus crucial in microemulsion research. Their determination has been attempted by various indirect methods, among which the light and neutron scattering techniques seemed to be promising. In particular, the neutron spin echo (NSE) in combination with small-angle neutron scattering (SANS) are powerful methods in which, in principle, not only the static characteristics of the droplets (the droplet radii, the thickness of the surfactant layer, the polydispersity in the sample) can be determined but also the thermal shape fluctuations of these (expectably flexible) objects can be studied. The first NSE studies are related to 1987 [1]. In those experiments a peak in the $q$ (the wave-vector transfer at the scattering) dependence of the effective diffusion constant $D_{eff}$ of the droplets was observed. The peak was attributed to the shape fluctuations of the droplets and from its height the bending elasticity $\kappa$ has been determined on the basis of the theory by Milner and Safran [2]. This coefficient was estimated to be ~ 5 $k_BT$, the value which differs markedly (by a factor from 2 to 10) from the values found using other methods (for more discussion see Refs. [3 - 7]). In the latter papers we expressed the opinion that the main reason for these discrepancies consists in an improper account for the shape fluctuations in the interpretation of the experimental data, and proposed a modification of the previous approach. Moreover, we have shown that even within the original theory [2] the interpretation of the NSE experiments was not carried out correctly [4]. We have found both by analytical estimations and numerical calculations that, instead of the values ~ 5 $k_BT$, the experiment [1] yields for $\kappa$ a value almost 14 times smaller [4], if one tries to fit the observed height of the peak in $D_{eff}(q)$. The difference from the theory [2] was that we considered different scattering length densities of the microemulsion components, the decay rates of the fluctuations were calculated for incompressible membranes, and higher modes (not only the lowest $l = 2$ mode) of the fluctuations have been included into the consideration. We would like to underline that none of the existing theories is able to explain the height of the observed peak in $D_{\text{eff}}(q)$: for the sets of droplet parameters encountered in the literature it is much lower than the current theories predict. Possible explanations of this contradiction are mentioned in Ref. [4] (that the droplet fluctuations are in fact larger than it is thought – in this case however the current theories are not valid), or in Ref. [3] (that the energy dissipation in the surfactant layer should be taken into account).

___

[*]E-mail address: lisy@upjs.sk

With time going $\kappa$ extracted from the experiments by Farago and coworkers has lowered up to a value ~ 3.6 $k_BT$ in 2001 [8]. Our criticism to that work in Ref. [9], particularly concerning the too high value of $\kappa$ determined there, was named as completely irrelevant [10]. This did not hinder B. Farago in the same 2001 year to become a co-author of the paper [11] in which $\kappa \sim 1$ $k_BT$ has been found, in a sharp contrast with his almost 15 years' conviction. It is a place here to cite the recent paper by Hellweg *et al.* [12]: "… However, in early inelastic experiments with neutron spin-echo techniques, Farago and coworkers found bending constants values that were much larger than those deduced from static scattering techniques [13]. We showed recently that this discrepancy was due to difficulties in the analysis of the neutron inelastic scattering curves, which are noisy; the combination of neutron and light scattering helped to circumvent the difficulty, reducing the number of fitting parameters. After such an analysis, all the moduli determinations were consistent with each other [14]." Note: the cited works [14] are dated 1998 and 1999 and not signed by B. Farago. We suspect that Farago and coworkers, due to noisy experimental curves, so many years produced so unrealistic values for the bending moduli.

In our opinion also the work [14] and the more recent investigations [11, 15] by Hellweg and co-authors contain serious difficulties and should be revised. Our objections to these works have been expressed in Refs. [16, 17]; similar arguments are raised in [9]. Our comments in [16] had no reply. The responses in [10] (to [9]) and [12] (to [17]) under no circumstances can be considered as a final resolution of the existing problems in the description of SANS and NSE scattering experiments on droplet microemulsions. They are so full of incorrect statements, even untruths, that we are forced again to point out evident defects of those works: in the methodical and physical approach to the solved problems, as well as in a wider context.[**]

Our objections to the discussed works (mainly to the most recent experimental study [11]) and the corresponding responses are as follows (not in the order of importance):

**1) *Objection*:** The particle form factor for monodisperse thin shells, Eq. (30) in [11], is not correct: instead of the factor ~ $(\rho_{\text{int}} - \rho_{\text{ext}})^2$ before the square brackets there should be $(\rho_{\text{shell}} - \rho_{\text{ext}})^2$. Here $\rho_{\text{shell}}$, $\rho_{\text{int}}$, and $\rho_{\text{ext}}$ refer to the surfactant layer, interior of the droplet, and the surrounding fluid, respectively.

*Response*: The authors agree with us. However, they claim that there was only an unfortunate misprint in Ref. [11]. As will be seen below, "misprints" often appear in their works but they assure the reader that the experimental data were treated using correct formulae.

---

[**] We would also like to mention a strange behavior of Colloids and Surfaces A (CSA) that we encountered trying to publish our Comment [17]. The Comment was sent to CSA on October 22, 2001. Then we had no information about the refereeing of the manuscript until April 17, 2002, when we received (after three our inquiries) a one sentence E-mail informing us that CSA is "waiting for a second review". On November 1 and 24 we again sent letters to the Journal, both with no answer. Our last letter to the editor was dated February 1, 2003. Finally, this short (less than 2 pp.) Comment was accepted on February 11, 2003 together with Reply by Hellweg *et al.* on more than 5 pp. One could wonder why this process took so long if, as indicated on the published papers, the Reply was received by CSA at the same day as Comment (November 5, 2001).

**2) *Objection*:** Equation (3) in Ref. [11] does not take into account droplet fluctuations in the shape and is thus incorrect. It also contradicts the intermediate scattering function $I(q,t)$, Eq. (6). We can add that Eq. (3) is in contradiction with the aim of the paper [11], which was to study the shape fluctuations of microemulsion droplets.

*Response*: The authors agree that using Eq. (3) they indeed neglect the other types of shapes (except that of a spherical shell) produced by thermal fluctuations. They add for some reason that they do not, however, neglect the differences between scattering length densities, and account for finite shell thickness. They also express the meaning that the fluctuations produce radius variations in the first place, and by use of a distribution in radii they "account for the resulting polydispersity, which is intrinsic to the system, due to thermal fluctuations, and facilitated kinetically by collisions among the droplets".

Hellweg *et al.* also insist in Ref. [12] that the thin shell approximation is satisfactory in the description of the data. If so, in their treatment of SANS the formfactor of the static scattering would be simply $f_0(qR) = j_0^2(qR)$. We have however analyzed in detail in Ref. [3] that the correct interpretation of SANS requires taking the fluctuations into account. The authors [11] implicitly assume very large $\kappa$ when treating SANS. Naturally, a combined analysis of such a treatment of SANS and the different consideration of the inelastic experiments (where the fluctuations appear) is doubtful.

**3) *Objection*:** The intermediate scattering function $I(q,t)$, Eq. (6) in [11], is incorrect. Its time-independent part is determined by the function (Eq. (7) in [11]) but with $l > 1$ instead of $l > 2$ in the sum):

$$f_0(qR) = j_0^2(qR) + j_0(qR)[(4 - qR)j_0(qR) - 2qRj_1(qR)]\sum_{l \geq 2}\frac{2l+1}{4\pi}\langle u_{l0}^2\rangle, \qquad (1)$$

where $u_{l0}$ are fluctuation amplitudes of the radius. Even if the constraint on the droplet volume (see below) is not considered, this equation should be changed: instead of the factor before the sum there should be

$$j_0(qR)[(2 - q^2R^2)j_0(qR) - 2qRj_1(qR)] \ . \qquad (2)$$

This result from Ref. [3] was later confirmed by Farago and Gradzielski in Ref. [8] (without a reference to our earlier paper).

*Response*: This response is typical for the criticized authors. They (Ref. [12]) give the corrected expression, which was however not used in their original work [11]. In fact, they implicitly agree with us. In Ref. [11] they indeed used the incorrect expression for $f_0$ (independently of whether they took into account the conservation of the droplet volume or not). Or was it just another unfortunate misprint? Their humor concerning the identical notation $l > 1$ and $l \geq 2$ is needless, because they really have $l > 2$ in Eq. (7), as everybody can check. The authors also note that the second term of $f_0(qR)$ is missing in Ref. [2] by Milner and Safran, and (again) there is a misprint in Ref. [13] by Farago *et al*.

**4) *Objection*:** So, $f_0(qR)$ has been used incorrectly. We added to the objection 2) that even correcting $f_0(qR)$ as shown above, it is still incorrect since it does not account for all the second-order terms in $u_{l0}$. The correct result for $f_0(qR)$ in the limit $d \to 0$ and the shell contrast ($\rho_{int} = \rho_{ext}$) differs from the above expression by the factor before the sum, which should be simpler, namely [3, 17]

$$-j_0^2(qR)(2 + q^2R^2). \qquad (3)$$

The difference arises due to the term ~ <$u_{00}$> as a consequence of the conservation of the droplet volume. We argued that calculating any physical quantity (if the droplet volume conserves) to the second order in $u_{lm}$, the terms ~ <$u_{00}$> cannot be neglected. As the simplest examples we showed that for the volume of the droplet we obtain $V = 4\pi R^3/3$ (omitting the $u_{00}$ term, the volume would be $(4/3)\pi R^3 [1 + (3/4\pi)\sum_{l>1,m}|u_{lm}|^2]$), the correct expression for the surface area is $A = 4\pi R^2 + (R^2/2)\sum_{l>1,m}(l-1)(l+2)|u_{lm}|^2$ instead of the incorrect $4\pi R^2 + (R^2/2)\sum_{l>1,m}(l^2+l+2)|u_{lm}|^2$, *etc*. The formulae used in Ref. [11] (and other papers by the authors) implicitly assume the volume conservation since they use the results of Ref. [2] by Milner and Safran, where this constraint is explicitly introduced.

*Response*: There is no real response to this objection. The authors contend that they use small fluctuation theory only for the $l > 1$ modes and treat the $l = 0$ mode differently by averaging all the expressions by a polydispersity function. There is no reason to discuss more this obscure argument. The patient reader can immediately see that such an approach leads to incorrect results. Consider, for instance, a simple example as that above for the surface area of the droplets. Let $\varepsilon$ be the (small) polydispersity in the sample and the radii are distributed according to Schultz or Gauss distribution (the latter follows from the microemulsion thermodynamics and is thus more natural to be used). Then for the radius one has the average over the distribution in radii, neglecting much smaller terms,

$$\langle R^2 \rangle \approx \langle R \rangle^2 (1+\varepsilon). \tag{4}$$

The correct result for the mean area is thus

$$\langle A \rangle \approx 4\pi \langle R \rangle^2 (1+\varepsilon) + \frac{1}{2}\langle R \rangle^2 (1+\varepsilon) \sum_{l>1,m} |u_{lm}|^2 (l-1)(l+2), \tag{5}$$

where the second term simplifies to $(1/2)\langle R \rangle^2 \sum_{l>1,m}(l-1)(l+2)|u_{lm}|^2$. This term would be, according to Hellweg *et al.*, $(1/2)\langle R \rangle^2 \sum_{l>1,m}(l^2+l+2)|u_{lm}|^2$, i.e. for the $l = 2$ mode two times larger than the correct one. The full average incorporates also the averaging over the fluctuations $u_{lm}$. The corresponding formula for small volume fractions of the droplets is [3]

$$\langle |u_{lm}|^2 \rangle = \langle u_{l0}^2 \rangle = \left\{ (l-1)(l+2)\left[ \frac{\kappa}{k_B T} l(l+1) - \frac{1}{8\pi\varepsilon} \right] \right\}^{-1}. \tag{6}$$

The same approach should be applied averaging any other quantity, including the static or dynamic structure factors of the droplets.

We emphasize that it is not true that as we have used a different definition of the radius from that in Ref. [11]. We use exactly the radius of a sphere with the same volume as the droplet. Expecting such an argument, we have explicitly pointed it out in Refs. [17, 9]. In spite of this Hellweg *et al*. insist on our definition of $R$ as an average radius [12].

They also comment on our sentence from Ref. [3] where we say that the samples always contain droplets of different radii: …"This leaves impression that they do not recognize that the polydispersity in radius arises from the fluctuations $u_0$ in microemulsion systems. This seems to overcount the intrinsic polydispersity effects … and could be the main reason for the discrepancy…", p. 260 [12] (underlined by VL). We believe that it is the misleading approach of Hellweg *et al*. that has been clearly shown above. The distribution of the droplets in radii follows from the thermodynamics of the droplet formation and even in the cases when the fluctuations are fully negligible, the sample really contains droplets with different radii. Our impression is that the authors [12] did not discussed this question with one another, since at least two of them agree with us, see Ref. [10]: "… *a priori* one could suppose that in fact the microemulsions are very monodisperse and the apparent polydispersity is only due to large shape fluctuations. This idea came up already at the very first experiments and was incompatible with the experimental data."

**5) *Objection*:** The used expression for the decay rates $\Gamma_l$ (which are not the relaxation times as they are called in Ref. [12]) is improper. The authors use our formula for the relaxation times of the droplet fluctuations [18][***] (citing the later work [20]), which, however, corresponds rather to highly compressible monolayers, while the monolayers of microemulsion droplets are commonly thought to be almost incompressible.

*Response*: The authors after some reasoning agree with us in this point. They add that when using the correct expression, the difference in the relaxation times is small (never larger than a few percent); their use of the incorrect expression in Ref. [11] thus cannot explain the change of elastic constants found by us.

Hellweg *et al.* can claim it only because they did not carry out more careful calculations. They consider only the $l = 2$ mode; the influence of the higher modes is however essential in the calculation of the scattering functions. The corresponding analysis can be found in our paper [4]. In some cases the $l = 2$ mode contributes only about one half to the total sums entering the scattering function [4]. For higher $l$, the use of the correct relaxation times brings significant changes in the scattering functions. We stress that considering the only $l = 2$ mode cannot lead to reliable extraction of the bending moduli from experiments.

There are more things to discuss in Ref. [12], e.g. our inclusion in the scattering function the corrections due to the finite thickness of the surfactant layer. We explained it already in Ref. [3] so we shall not dwell on this question. But we are forced to give one more objection not present in our previous papers.

**6) *Objection*:** It concerns the way of discussion used by the criticized authors. It is partially demonstrated above (e.g. point 3)), but well seen by the following example. In Ref. [10] they "show" that the expression they use for $P_{\text{stat\_corr}}(q)$ in [8] is the same one as our formula derived earlier in Ref. [3]. How they are doing this: They take our Eq. (5) in [3] and substitute there Eqs. (3) and (6). Then, according to Ref. [10]: "…it does not take long to realize that this equation is *strictly* identical to Eq. (14) in our paper" (Ref. [8], or Eq. (2) in Ref. [10]). They do not see from where the new (proposed by us [9]) Eq. (1) [10] is coming from. The explanation is unfavorable for Farago and Gradzielski: taking our Eq. (5) [3], they do not use its complete form corresponding to $P_{\text{stat\_corr}}(q)$. They have taken only a part of this equation to obtain the result used in Ref. [8]: of course, the incorrect result. Incorrect expressions for the static correlators are used in all their papers. As already mentioned above, they agree with this (except the case considered in this Objection) but claim (with exclamation mark) that in the data treatment the correct equations were used.

They also declare that we suggest to forget about NSE data which directly (again with "!") measures the fluctuation dynamics and rather fit all parameters to the set of SANS data. This exclamation is fully wrong since such a suggestion cannot be found in our papers. We criticized the approach of Ref. [8] owing to the inconsistent determination of the droplet parameters, but did not suggest forgetting about NSE.

---

[***] The authors know our results [18]. After the appearance of the paper [18] S. Safran [personal communication, October 8, 1991] discussed with us their correctness, particularly the compatibility of our decay rates with the scaling $\sim R_0^{-3}$ observed in the experiments [1]. The decay rates are compatible with this scaling, as seen from Refs. [11, 12] (we have shown it already in the subsequent paper [19] where also the bending elasticity was extracted from the experiments [1]; at that time we estimated it to be almost three times smaller than in the original work).

One can also find in Ref. [10] that we used "*13* (!)" data points in a limited *q* range to obtain the droplet parameters. The answer is very simple. In many cases the only point is enough to judge about the description of the experiments (e.g. the description of the NSE experiments [1, 13] that yields about one order larger height of the peak in $D_{eff}$ than the observed one). Moreover, as we write in Ref. [3] (p. 4050), our calculations should be considered only as estimations of the system parameters. We have proposed a method of interpretation, not the numerical values of the parameters. It is however true that, substituting the parameters determined by the criticized authors in our equations, the found discrepancy with the experiments is so marked that it prejudices the correctness of these parameters.

In conclusion, according to Ref. [12], the authors explain there how shape fluctuations and polydispersity are coupled in microemulsion systems in which the total surface area is a fixed quantity. We cite: "Other interpretations by Lisy and coworkers do not agree with ours on this particular point. This is probably the reason why they obtain unrealistic values of the bending elastic constants, whereas the description that we use leads to excellent agreements with independent determinations of these constants." In this note we have shown that the authors were not able to support their description. Quite on the contrary, it was easy to reveal that both from the theoretical and methodical points of view their approach is incorrect. We have demonstrated that the interpretation of the experiments [11] and all other cited experiments on droplet microemulsions by Hellweg *et al.* is flawed. We suspect that they apply erroneous description of the experiments and provide us thus with unreliable values of important parameters of microemulsions, particularly, the bending elastic constants. We also suspect that they use incorrect and not clear methods of discussion. These authors agreed with a part of our arguments. In some points they do not agree with us, but their responses are inessential. Some of the named authors have already recognized [12] that the values of the constants $\kappa$ and $\bar{\kappa}$, as they have extracted them from earlier experiments, must be significantly changed. We believe that their other experiments will be reconsidered as well.

_____________________________